\documentclass[prd,superscriptaddress,twocolumn,preprintnumbers,amsmath,amssymb]{revtex4-2}

\usepackage[breaklinks,colorlinks=true]{hyperref}
\usepackage{graphicx}
\usepackage{mathtools}
\usepackage{slashed}
\usepackage[caption=false]{subfig}

\usepackage[normalem]{ulem}

\begin{document}
\title{Emergent spacetime from a Berry-inspired dynamical gauge field coupled to electromagnetism}
\author{Patrick Copinger}
\email{copinger0@gate.sinica.edu.tw}
\affiliation{Institute of Physics, Academia Sinica, Taipei 11529, Taiwan}
\author{Pablo Morales}
\email{pablo$_$morales@araya.org}
\affiliation{Research Division, Araya Inc., Tokyo 107-6019, Japan}
\affiliation{Centre for Complexity Science, Imperial College London, London SW7 2AZ, UK}


\begin{abstract}
Motivated by the fermionic Berry's phase in momentum space, we study a local Abelian phase in momentum space coupled to electromagnetism, for complex scalars in the phase-space worldline formalism. The interaction of both Abelian fields is shown to give rise to a momentum gauge dependent emergent spacetime. As a concrete example, we further study classical solutions of the Berry-inspired gauge field that lead to an emergent Newtonian gravity with gravitational potential predicted by coupled Coulomb fields both in configuration and momentum spaces. Noncommutative aspects of the theory are also provided.
\end{abstract}

\maketitle

\section{Introduction}
Berry's phase~\cite{doi:10.1098/rspa.1987.0131} is a ubiquitous concept in condensed matter: It underpins the Hall current and conductivity~\cite{Thouless:1982zz,*PhysRevLett.75.1348,*PhysRevB.53.7010}, the anomalous Hall effect~\cite{PhysRev.95.1154,*doi:10.1126/science.1058161,*PhysRevLett.88.207208,doi:10.1126/science.1089408}, and spin transport~\cite{doi:10.1126/science.1087128,*PhysRevLett.92.126603}, to name a few. The phase exists in Weyl/Dirac semimetals at Weyl nodes in quasicrystal momentum space, and gives rise to the chiral (Adler-Bell-Jackiw~\cite{PhysRev.177.2426,*ref1}) anomaly~\cite{PhysRevX.5.031023,*doi:10.1126/science.aac6089}. A remarkable feature is the classical emergence of the anomaly in a Louiville theorem for Weyl fermions by the way of an incompressible phase-space~\cite{PhysRevLett.109.162001}. Remarkable because, despite the Berry phase's prevalence, the phase's coupling to electromagnetism gives rise to ``exotic"~\cite{DUVAL2000284,*Horvathy:2010wv} noncanonical--equivalently noncommutative in the quantum picture--d.o.f. in phase-space~\cite{C_Duval_2001,*PhysRevD.69.127701,doi:10.1142/S0217984906010573}, making possible the anomaly. 

Noncommutative theories have been extensively studied; see, e.g., reviews by~\cite{RevModPhys.73.977,*SZABO2003207}. A theory was first formulated by Snyder in an efforts to combat the UV divergences plaguing quantum electrodynamics as a natural regularization~\cite{PhysRev.71.38}. Similar encompassing noncommutative theories in momentum space have ensued under doubly, or deformed, special relativity~\cite{AMELINOCAMELIA2001255,*KOWALSKIGLIKMAN2001391,*doi:10.1142/S0218271802001330} including those of $\kappa$-Minkowski~\cite{LUKIERSKI199590}, and Magueijo-Smolin~\cite{PhysRevD.67.044017}. noncommutative theories can arise in systems with electrons subject to a strong magnetic field~\cite{Szabo:2009tn}, such as has been applied for the Peierls substitution~\cite{peierls1933theorie}, and they naturally occur in string theories~\cite{Alain_Connes_1998}. The exotic Galilei symmetry has also been studied for anyons in~\cite{Cortes:1995wa,*
delOlmo:2005md}. A feature that goes hand-in-hand with noncommutativity is an emergent spacetime geometry. This has been conceived for (complex) scalars~\cite{RIVELLES2003191} and gauge field theories~\cite{doi:10.1142/S0217732306021682,*doi:10.1142/S0217751X0904587X} using a Sieberg-Witten map~\cite{Nathan_Seiberg_1999}, as well as matrix theories~\cite{Harold_Steinacker_2007,*STEINACKER20091}. Furthermore, the connection to curved momentum space has been put into the more formal mathematical language of Hopf algebras~\cite{majid_1995}. It is thus intuitive that an emergent geometry should be present in deformed phase-space courtesy of Berry's phase; indeed hints to such a relationship were found in~\cite{PhysRevD.91.025004}, where the fermionic covariant Berry's phase was likened to a spin connection.

We explore emergent spacetime geometries from a Berry phase inspired local momentum phase. By ``emergence" here we mean features of gravity, or a spacetime metric, can arise (or emerge) without a priori introducing gravitational d.o.f. Whereas monopoles (magnetic charges) are understood in momentum space from Berry's phase, locality makes possible electric charges in momentum space. By way of a quantum field theory defined in momentum space the local symmetry was introduced as a ``momentum gauge" in~\cite{Guendelman:2022gue}, and has since been derived from a Kaluza-Klein reduction in curved momentum space~\cite{Guendelman:2022ruu}. In~\cite{Guendelman:2022gue} it was demonstrated through a dynamical version of Born’s reciprocity theory how a momentum gauge naturally follows; it was furthermore shown how non-commutativity arises. Here we show how the momentum gauge shares a similar form to the adiabatic Berry phase in momentum space. Also, we introduce a coupling to electromagnetism. This is important. For without it in our first-quantized setting, noncommutativity and the resulting emergent spacetime cease. Furthermore, we go on to argue that electric charges in the Berry-inspired gauge coupled to an electromagnetic charge give rise to a weak Newtonian gravity with the product of field strengths in both configuration and momentum spaces resembling a Newtonian potential.
\vspace{0.5em}

\section{Fermionic Berry Phase Motivations}
Let us motivate our discussion of a dynamical gauge in momentum space by first introducing its occurrence as an Abelian Berry's phase for fermions after employing the adiabatic theorem. Berry's phase is easily captured in a first-quantized setting in phase-space; furthermore a manifestly Lorentz invariant description is desirable. Therefore we make use of a phase-space worldline construction~\cite{MIGDAL1986594}, (c.f., applications to Snyder spacetime in~\cite{Bonezzi_2012,*PhysRevD.98.065010}). We review relevant discussions as outlined in~\cite{PhysRevD.105.116014}; see also~\cite{PhysRevLett.109.162001,PhysRevD.88.045012,*PhysRevD.89.094003,*Dwivedi_2014}. To begin let us write down the (n+1)-dimensional massive fermionic Green's function for quantum electrodynamics using a Minkowski flat-space metric (in Greek letter indices) with mostly-plus signature in Cartesian coordinates as
\begin{equation}\label{eq:S_f}
    S(x_f,x_i)=i\int^\infty_0dT\int \mathcal{D}x\mathcal{D}p\mathcal{D}A\,\mathcal{P}e^{\frac{i}{\hbar}[S_\textrm{F}+S_A]}\,.
\end{equation}
Here the fermionic worldline action reads
\begin{equation}\label{eq:fermion_action}
    S_\textrm{F}=\int^1_0d\tau[\dot{x}_\mu p^\mu+qA_\mu \dot{x}^\mu -T(\slashed{p}+mc)]\,,
\end{equation}
with Maxwell action, $S_A=-(4\mu_{0}c)^{-1}\int d^{4}x F_{\mu\nu}F^{\mu\nu}$, for the dynamical gauge field~\cite{KARANIKAS1992176}. Note that we have excluded the normalizing determinant factor under the gauge integration in a ``quenched approximation"~\cite{PhysRevD.100.105020}. The determinant will not affect our semiclassical picture employed for scalars below. There is a path-ordering in propertime acting on the fermionic indices. We have also taken the liberty of absorbing $T$ into $\tau$, where now the Schwinger propertime acts a Lagrange multiplier. Note that we have made use of SI units here and throughout. Finally, one has for path integral measure boundary conditions, $x(0)=x_i$ and $x(1)=x_f$. 

Berry's phase manifests from a similarity transform of the Hamiltonian; in our context this occurs via a Hamiltonian gauge-transformation of $\slashed{p}\rightarrow u^{-1}\slashed{p}u+i\hbar u^{-1}\dot{u}$~\cite{PhysRevD.105.116014}, where $u$ may be chosen to take $\slashed{p}$ diagonal. Then under the adiabatic theorem level-crossing terms in $u^{-1}\dot{u}$ may be neglected, leading to a connection with non-vanishing curvature; such a process facilitates the chiral anomaly for Weyl fermions~\cite{PhysRevLett.109.162001}. In a similar way, on the worldline in (2+1)-dimensions off-diagonal terms may be neglected leading to Abelian Berry's phases: $\mathrm{diag}[(u^{-1}\partial_\mu^p u )^{11},(u^{-1}\partial_\mu^p u )^{22}]$; $\partial_\mu^p \coloneqq \partial/\partial p^\mu$. Under the adiabatic theorem, a large separation in eigenvalues is required. Here, on the worldline, this amounts to a large mass approximation since the eigenvalues of $\slashed{p}$ are $\pm \sqrt{p^2}=\pm m^2$ on the mass shell; see~\cite{PhysRevD.105.116014} for further details. Note, this differs from the conventional Hamiltonian leading instead to an energy gap.
The Berry phases then supplement the previous worldline action, Eq.~\eqref{eq:fermion_action}, with a force dependent term; e.g., for the ``11" matrix index element, upon decoupling of the path-ordering, the corresponding Lagrangian would now read
\begin{equation}\label{eq:Berry_Lag}
    \dot{x}_\mu p^\mu+qA_\mu \dot{x}^\mu 
    +\hbar (u^{-1}\partial_\mu^p u)^{11}\dot{p}^\mu-T[(u^{-1}\slashed{p}u)^{11}+mc]\,.
\end{equation}
Note that in the worldline representation, the Lagrangian after a similarity transformation retains Lorentz invariance.
The introduction of the Berry phase term spoils the canonical Poisson bracket structure~\cite{C_Duval_2001,*PhysRevD.69.127701,doi:10.1142/S0217984906010573}, giving rise to noncanonical d.o.f. The conserved phase-space volume is also augmented~\cite{PhysRevLett.95.137204,*PhysRevLett.96.099701}. A key feature of the above modified Lagrangian is the necessity of a U$_x(1)$ electromagnetic coupling~\cite{PhysRevLett.109.162001}, for without it one may simply absorb the Berry phase (after integration by parts) into a coordinate redefinition as $x_\mu\rightarrow x_\mu+\hbar (u^{-1}\partial_\mu^p u)^{11}$ rendering physical contributions from the Berry phase trivial.

Motivated by the appearance of a momentum space dependent Berry phase, in this work we explore an extended local Abelian U$_p(1)$ symmetry in momentum space, taking for example $(u^{-1}\partial_\mu^p u)^{11}\rightarrow B_\mu$, where now the momentum gauge connection $B_\mu(p)$ may represent an arbitrary function. We hypothesize the local symmetry because: 1. The pure gauge transformation should already present; a gauge transformation, with $u\in$ U$_p(1)$, of the Berry phase will give no observable effects to the accumulated phase. 2. The local momentum gauge symmetry is already a symmetry for the free theory without electromagnetism. This is most simply demonstrated, like above, with a coordinate shift: $x_\mu\rightarrow x_\mu+bB_\mu$ for coupling $b$. Non-triviality appears from an interaction with electromagnetism.

The introduction of a U$_p(1)$ local symmetry begs the question: Why has it not been observed? We argue this is because it appears, due to its noncommutative nature, as an emergent spacetime. We clarify this connection by examining the simplest scenario with a U$_p(1)$ and U$_x(1)$ interaction, one of two well-separated dissimilar dual-charged particles, demonstrating the appearance of a Newtonian potential.

\section{Complex Scalars with Dynamic Gauge in Momentum Space}
The simplest physical setting in which a local U$_p(1)$ phase in momentum space would have physical effects is one for a complex scalar coupled to electromagnetism. Let us denote for Abelian gauge in momentum space, $B_\mu$, a curvature of $S_{\mu\nu}=\partial^p_\mu B_\nu - \partial_\nu^p B_\mu$. Now we postulate, based on the above Berry phase motivations, the most general propagator with dynamical and local momentum gauge, written in a worldline representation for complex scalars as
\begin{equation}\label{eq:propagator}
    G(x_f,x_i)=\int^\infty_0\hspace{-0.6em} dT\int \hspace{-0.2em}\mathcal{D}x\mathcal{D}p\mathcal{D}A\mathcal{D}B\sqrt{-\omega}\,e^{\frac{i}{\hbar}[S_\text{W}+S_A+S_B]}\,,
\end{equation}
with worldline action given by 
\begin{equation}\label{eq:w_action}
    S_\text{W}=\int_{0}^{1}d\tau[\dot{x}_\mu p^\mu+qA_\mu\dot{x}^\mu+bB_\mu\dot{p}^\mu-T (p^{2}+ m^{2}c^{2})]\,.
\end{equation}
Analogous to $S_A$, we have introduced a similar Maxwell-like kinetic term for the Berry-inspired gauge given by $S_B =-(c/4s)\int d^{4}pS_{\mu\nu}S^{\mu\nu}$.
We work in (3+1)-dimensions. We have placed a factor of $c$ that arises from energy, $cp^0$, and in analogy to the vacuum permeability we have introduced a factor $s$. The dimension of $[bB_\mu]$ is given as a length. Different qualitatively from Eq.~\eqref{eq:S_f}, we have included the factor $\sqrt{-\omega}$, with $\omega\coloneqq\mathrm{det}\omega_{\mu\nu}$, for an invariant phase-space measure for the noncanonical d.o.f.~\cite{Jackiw:1993in,*MIGNEMI20161714}--an explicit form for $\omega$ will be given later. 
We stress that here all indices are flat; we will, however, explore an emergent metric shortly. 

Our aim in this work is to explore the classical and interacting physics of Eq.~\eqref{eq:w_action}. While the placement of the gauge in momentum space is similar to Berry's phase in Eq.~\eqref{eq:Berry_Lag}, the importance of integrating over $B_\mu$ stems from the fact that now $B_\mu$ [as a classical solution of some point-like particle] may be treated as a background field [of another point-like particle]. Indeed, writing $q\int d\tau A_\mu\dot{x}^\mu=c^{-1} \int d^{4}y\,j_\mu A^\mu(y)$, one may find for a point-like particle an electromagnetic current given by $j^\mu(y) =qc\int d\tau\dot{x}^\mu \delta^{4}(y-x(\tau))$ that is conserved, $\partial_\mu j^\mu =0$, with charge $\int d^{3}y\,j^{0}(y)=qc$~\cite{Balasin:2014dma}. In the same way, one may find an analogous current in momentum space from $b\int d\tau B_\mu \dot{p}^\mu =c\int d^{4}qj_{p}^\mu B_\mu(q)$ with corresponding charge as
\begin{equation}\label{eq:j_p}
    j_{p}^\mu(q) = \frac{b}{c}\int d\tau\dot{p}^\mu\delta^{4}(q-p(\tau))\,,\quad \int d^{3}qj_{p}^{0}(q)=\frac{b}{c}\,;
\end{equation}
it too is conserved, $\partial_\mu^p j_{p}^\mu=0$. Such a conserved current in momentum space was first envisioned in~\cite{Guendelman:2022gue}. One may push the point-like particle analogy further by considering the classical solution in momentum space: Like the inhomogeneous Maxwell equation, i.e., $\partial_\mu F^{\mu\nu} =-\mu_{0}j^{\nu}$, one has
\begin{equation}\label{eq:maxwell}
    \partial_\mu^p S^{\mu\nu}=-sj_{p}^{\nu}\,,
\end{equation}
which admits a Coulomb-like solution in momentum space~\cite{Guendelman:2022gue}. 

It has been established that noncommutative systems may be interpreted as a curvature in momentum space~\cite{majid_1995}. However, we would like to explore an emergent curved space in a conventional coordinate space representation. It is anticipated this may be achievable by integrating out the momenta. Since we are interested in augmentations to the classical picture, we employ a semiclassical technique. In the absence of a momentum gauge but in curved coordinate space, such a technique can reproduce the configuration-space action from its phase-space description~\cite{PAVSIC1987327}. Let us begin by writing down the equations of motion in $x^\mu$ and $p^\mu$:
\begin{equation}\label{eq:eoms}
    \dot{x}^\mu=2T p^\mu-bS^{\mu\nu}\dot{p}_\nu\,,\quad \dot{p}^\mu=qF^{\mu\nu}\dot{x}_\nu\,.
\end{equation}
The above are a Lorentz covariant extension of the common Berry curvature modified dynamics of a Bloch electron in a solid, however for arbitrary curvature $S_{\mu\nu}$. Combining the two we have
\begin{equation}\label{eq:omega}
    p^\mu=\tfrac{1}{2T} \omega^{\mu\nu}\dot{x}_\nu \coloneqq \tfrac{1}{2T} (\eta^{\mu\nu}+S^{\mu}_{\hphantom{\mu}\sigma}F^{\sigma\nu})\dot{x}_\nu\,.
\end{equation}
Then taking solutions to the coupled Eq.~\eqref{eq:eoms} in $p^\mu$ as our classical solution, we may rewrite the worldline action, Eq.~\eqref{eq:w_action}, as
\begin{equation}\label{eq:curved_action}
    S_{\text{W}}=\int_{0}^{1}d\tau\bigl[\tfrac{1}{4T}\dot{x}^\mu g_{\mu\nu}\dot{x}^{\nu}+qA_\mu\dot{x}^\mu-T m^{2}c^{2}\bigr]\,,
\end{equation}
where we have suggestively written
\begin{equation}
    g_{\mu\nu}=\bigl[\eta-(qb)^{2}FS^{2}F+2qbF\partial^p B(\eta+qbSF)\bigr]_{\mu\nu}\,.
\end{equation}
In arriving at the above we have neglected surface terms after an integration by parts, $bB_{\mu}\dot{p}^{\mu}=-b\partial_{\nu}B_{\mu}\dot{p}^{\nu}p^{\mu}$, which lead to the same equations of motion in Eq.~\eqref{eq:eoms}. $B_\mu(p(x))$ and $S_{\mu\nu}$ are now functions of $x_\mu$. Contracted Lorentz indices are assumed where not explicitly written in matrix form, i.e., $[FS]_{\mu\nu}=F_{\mu}^{\hphantom{\mu}\sigma}S_{\sigma\nu}$. Note also that we do not treat fluctuations about the classical solutions in this analysis.

Let us treat the small electromagnetic coupling case keeping terms to $\mathcal{O}(q)$ in the metric as
\begin{equation}\label{eq:metric}
    g_{\mu\nu}\simeq\eta_{\mu\nu}+2qbF_{\mu}^{\hphantom{\mu}\sigma}\partial^p_\sigma B_\nu\,.
\end{equation}
Also since $-\mathrm{det}g_{\mu\nu}\simeq1+qb\eta^{\mu\nu}F_{\nu\sigma}S_{\hphantom{\sigma}\mu}^{\sigma}$, we find $\sqrt{-g}\simeq\sqrt{-\omega}$.
Therefore we can now interpret an emergent curved space system with $(g^{-1})^{\mu\nu}\eqqcolon g^{\mu\nu}$, and indices now represent those in curved space. The action in Eq.~\eqref{eq:curved_action} can now be interpreted as the usual curved complex scalar action to $\mathcal{O}(q)$; (note that then $qA_\mu\dot{x}^\mu$ may be treated with either flat or curved indices). And the invariant phase-space volume measure now becomes the conventional invariant curved space determinant in a sum over paths in coordinate space~\cite{BASTIANELLI2002372}. An interesting observation is that $\omega^{\mu\nu}$ closely resembles the emergent metric for the Maxwell Lagrangian in noncommutative spacetime with $\theta_{\mu\nu}\rightarrow S_{\mu\nu}$ found using an exact Sieberg-Witten map~\cite{doi:10.1142/S0217732306021682,*doi:10.1142/S0217751X0904587X}, for usual and constant noncommutative parameter in $[x_\mu,x_\nu]=i\theta_{\mu\nu}$.

A distinct feature of the induced metric, Eq.~\eqref{eq:metric}, is its gauge in momentum space dependence. Applying a gauge transformation, $B_\mu\rightarrow B_\mu+i\tfrac{\hbar}{b} u^{-1} \partial_\mu^p u$ for $u=e^{i\epsilon(p)}\in$ U$_p(1)$ to Eq.~\eqref{eq:metric} written symmetrically, we can see that additions to the metric, $q\hbar\{\partial_{\mu}^p\partial^p_{\sigma}  \epsilon F_{\hphantom{\sigma} \nu}^{\sigma}
+\partial_{\nu}^p\partial^p_{\sigma}  \epsilon F_{\hphantom{\sigma} \mu}^{\sigma}\}$, resemble additions from a gauge transformation in a linearized theory in general relativity (GR) as $\partial_\mu\xi_\nu+\partial_\nu\xi_\nu$ for infinitesimal diffeomorphism with Killing vector $\xi_\mu$, or a Lie derivative in linearized gravity. In this way we see that a momentum gauge dependence is a natural feature. 
We next explore a key example of an induced metric stemming from the backgrounds of a dual charged particle.
\vspace{0.5em}

\section{Dual U\boldmath{$_x(1)$} and U\boldmath{$_\text{p}(1)$} Coulomb Fields and Gravity}

\begin{figure}
\subfloat[]{%
\label{fig:setup}
\includegraphics[clip,width=0.9\columnwidth]{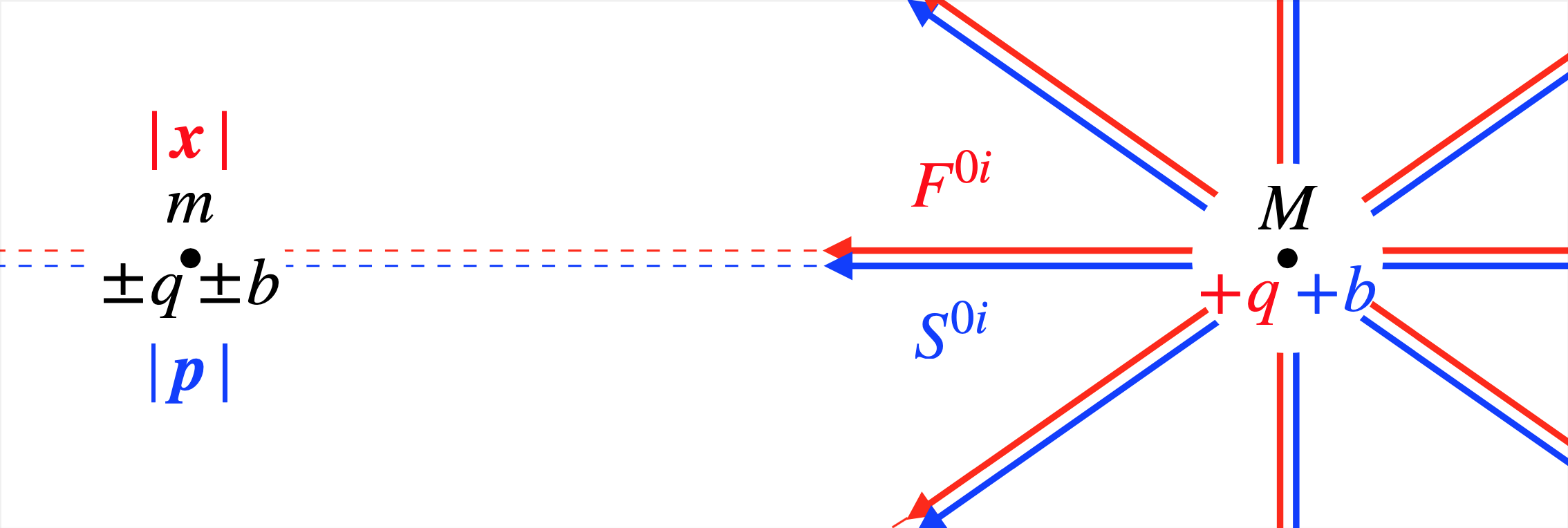}}\\
\subfloat[]{%
\label{fig:potential}
\includegraphics[clip,width=0.9\columnwidth]{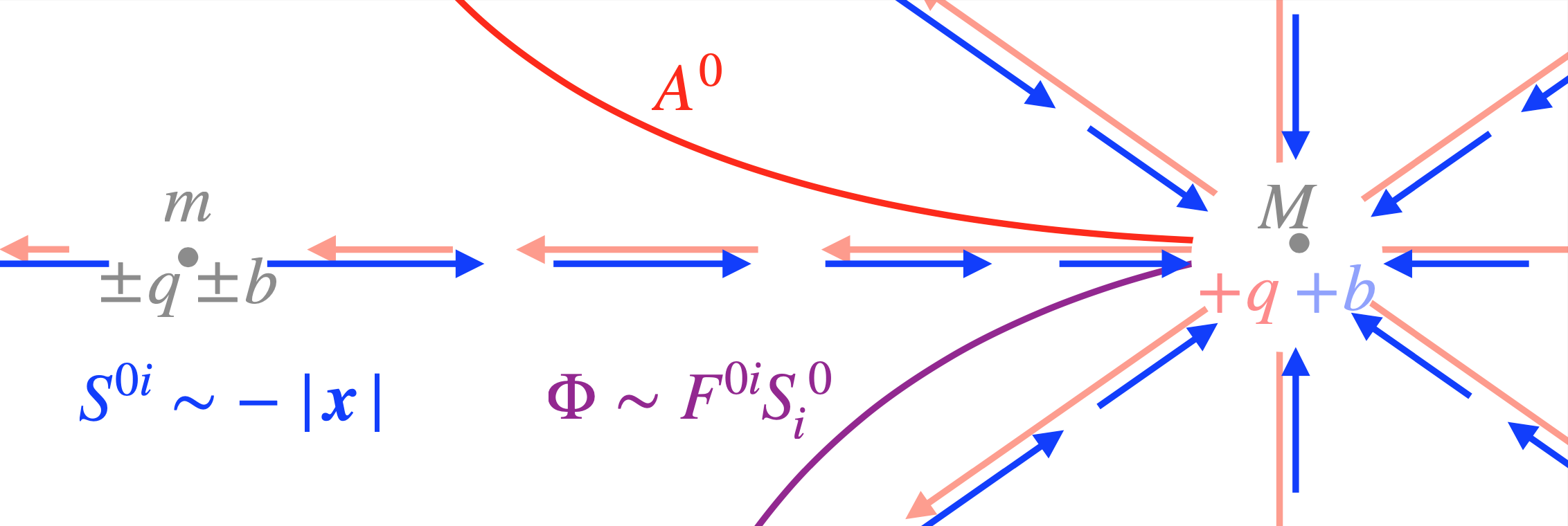}}
\caption{(a) Dual U$_p(1)$ and U$_x(1)$ background (generating $S^{0i}$ and $F^{0i}$ fields, Eq.~\eqref{eq:fields}, respectively) for particle of $M$ mass with sample $(+q,+b)$ charge. Test particle with $m$ mass is located at phase-space coordinate of $|\boldsymbol{x}|$ and $|\boldsymbol{p}|$. (b) Semiclassical picture of the dual background is shown in which $p_\mu(x)$ is found; this leads to $S^{0i}$ going as $-|\boldsymbol{x}|$. An emergent gravitational potential, $\Phi$, (shown figuratively as a function of $|\boldsymbol{x}|$) results from the product of both U$_p(1)$ and U$_x(1)$ fields given in Eq.~\eqref{eq:metric}. We also note that the original U$_x(1)$ potential, $A^0$, (also shown figuratively as a function of $|\boldsymbol{x}|$) is unmodified.}
\end{figure}

In the absence of the Berry-inspired gauge one could reason the simplest classical interacting scenario involving point-like particles would be one of a test charged particle in a U$_x(1)$ Coulomb potential. Therefore, let us extend the classical test particle picture to encompass the new U$_p(1)$ charge. In this picture one would have not only the Coulomb potential, but now also a Coulomb-like potential in momentum space~\cite{Guendelman:2022gue}. We take for the test particle an arbitrary mass, $m$, and for the particle generating the background field, a large mass, $M$, such that on-shell, $-p^\mu p_\mu=c^2M^2$, one may take for the energy
\begin{equation}\label{eq:mass_shell}
    \quad p^{0}=\pm\Bigl\{ Mc+\frac{\boldsymbol{p}^{2}}{2Mc}\Bigr\} +\mathcal{O}(M^{-2})\,.
\end{equation}
Next, let us consider for the background particle charges of $j^0=cq\delta^3(x)$ and additionally $j_p^0=(b/c)\delta^3(p)$, whose field solutions from the inhomogeneous Maxwell equation and Eq.~\eqref{eq:maxwell} are respectively
\begin{equation}\label{eq:fields}
    F^{i0}=\frac{-q}{4\pi\epsilon_0 c}\frac{\boldsymbol{x}}{|\boldsymbol{x}|^3}\,,\quad 
    S^{i0}= \frac{-sb}{4\pi c}\frac{\boldsymbol{p}}{|\boldsymbol{p}|^3}\,.
\end{equation}
We will also require a gauge selection; we choose a Coulomb gauge for each:
\begin{equation}\label{eq:gauges}
    A^0 = \frac{q}{4\pi\epsilon_0 c}\frac{1}{|\boldsymbol{x}|}\,,\quad 
    B^0= \frac{sb}{4\pi c}\frac{1}{|\boldsymbol{p}|}\,.
\end{equation}
See Fig.~\ref{fig:setup} for the background setup of a $(+q,+b)$ charged particle--(let us emphasize however that any $\pm q$ or $\pm b$ charge can be used).

Recall in Eq.~\eqref{eq:curved_action} to find an emergent metric we integrated out the momenta. To carry out this procedure, one needs to evaluate the equations of motion, Eq.~\eqref{eq:eoms}, for the background particle, and determine $p^\mu(x)$ and hence also $B_\mu(p(x))$. Note that we also make use of the on-shell equation of motion as dictated by the propertime, $T$, integral; according to Eq.~\eqref{eq:eoms}, one must always have $p^\mu p_\mu$ equal to a propertime-independent constant, and classical solutions will be dominant around the mass-shell criteria, Eq.~\eqref{eq:mass_shell}, for large $M$. Let us emphasize that although we are evaluating equations of motion, these are in-essence a self-interaction of the background particle onto itself. We also specify initial conditions for the background particle; we take that the particle was brought in from spatial infinity at rest such that $\boldsymbol{p}(0)=r(0)^{-1}=0$.

It is convenient to exploit the symmetry of our setup, and use a spherical coordinate system (in a coordinate basis). We use the conventional $x^\mu=(x^0,r\sin\theta\cos\phi,r\sin\theta\sin\phi,r\cos\theta)$, and spherical coordinate indices we write with Latin letters as $x^a=(x^0,r,\theta,\phi)$. Our transformation matrices read $e_a^{\hphantom{a}\mu}=\partial_{a}x^{\mu}$ and $e_\mu^{\hphantom{\mu}a}=\partial_\mu x^a$ such that $e_\mu^{\hphantom{\mu}a}e_a^{\hphantom{a}\nu}=\delta_\mu^\nu$ and likewise for contractions in Cartesian coordinates. Momenta are naturally lowered such that $p_a=e_a^{\hphantom{a}\mu}p_\mu$, and the flat space metric becomes $\eta_{ab}=\textrm{diag}(-1,1,r^2,r^2\sin^2\theta)$. Then one can write the equations of motion in Eq.~\eqref{eq:eoms} and Eq.~\eqref{eq:omega} in spherical coordinates as
$p^a=\tfrac{1}{2T}\omega^{ab}\dot{x}_b$, and $\dot{p}_{a}=\gamma_{ad}^{b}\dot{x}^{d}p_{b}+qF_{ab}\dot{x}^{b}$,
where we have for the connection with zero curvature, $\gamma_{ad}^{b}=\tfrac{1}{2}\eta^{bc}[\partial_{d}\eta_{ca}+\partial_{a}\eta_{cd}-\partial_{c}\eta_{ad}]$.

Exploiting the spherical symmetry we can evaluate $p(x)$ for the desired range of applicability. Let us first explicitly write out 
\begin{equation}
    p_{\theta}=\frac{r^2\dot{\theta}}{2T\rho}\,,\quad
    p_{\phi}=\frac{r^2\sin^2\theta\dot{\phi}}{2T\rho}\,,
\end{equation}
where we have
\begin{equation}
    \rho=1-\frac{1}{2T}\frac{sb^{2}q^{2}}{(4\pi c)^{2}\epsilon_{0}}\frac{\dot{r}}{|\boldsymbol{p}|^{3}r^{2}}\,.
\end{equation}
The key takeaway here is that even though $S^{ab}$ affects the magnitude of $p_\theta$ and $p_\phi$ through the $\rho$ factor, like the case without $S^{ab}$ they are still proportional to $\dot{\theta}$ and $\dot{\phi}$. Therefore, our setup has a similarity to the case of a stationary charged particle in the presence of a Coulomb potential for $r\gg 0$ in that the particle may only move in the radial direction, effectively reducing to (1+1)-dimensions. Hence we have that $p_\theta=p_\phi=0$, $|p_r|=|\boldsymbol{p}|$, and that $p_r$ is negative.

The time component of the Lorentz force equation, takes on a simple form with the selection of the Coulomb gauge in Eq.~\eqref{eq:gauges}, namely $\dot{p}^0=-q\dot{A}^0$. This differential equation can be readily evaluated using the initial conditions, as well as Eq.~\eqref{eq:mass_shell}, to find that
\begin{equation}\label{eq:p_r^2}
    p_r^{2}=\frac{Mq^{2}}{2\pi\epsilon_{0}r}\,.
\end{equation}
We have used the negative energy expression in Eq.~\eqref{eq:mass_shell}, which is needed for solutions with real momenta. 

Using the above arguments, along with the gauge selection in Eq.~\eqref{eq:gauges}, one can find that
\begin{equation}
    \partial_a^p B_b =S_{r0}=-\frac{sb}{2 c}\frac{\epsilon_0 r}{Mq^2}\,.
\end{equation}
Then we can determine that the induced metric, Eq.~\eqref{eq:metric}, becomes
\begin{equation}\label{eq:New_metric}
    g_{ab}=\textrm{diag}(-1+\Phi,1,r^2,r^2\sin^2\phi) \,,
\end{equation}
where we have that
\begin{equation}
    \Phi =\frac{1}{8\pi  r}\frac{sb^2}{Mc^2}\,.
\end{equation}
At this point we can now see the role of $s$ and $b$, and we fix their values such that $\Phi$ agrees with the Newtonian potential, or
\begin{equation}
    sb^{2}=16\pi M^{2}G\,.
\end{equation}
We have astonishingly predicted a new long range/low energy description of gravity, whereby the product of both Coulomb fields depict a gravitational potential. See Fig.~\ref{fig:potential} for a depiction of the combined fields living in coordinate space. The combination of both fields is such that neither $\pm q$ nor $\pm b$ charge dictates the sign of the potential; $q$ and $\epsilon_0$ cancel out and only the combination of $s b^2$ remains.

The metric given in Eq.~\eqref{eq:New_metric} before integrating out momenta was Berry-inspired gauge dependent. A different choice in Eq.~\eqref{eq:gauges} would have resulted in a different induced metric, e.g., supplying corrections to spatial components. However, let us again emphasize that ordinary gauge transformations in GR can accomplish the same thing; e.g., a synchronous gauge in perturbed gravity would also lead to corrections to spatial components instead of the Newtonian picture above~\cite{carroll_2019}. 


Let us last remark that a full description of a fictitious large mass particle would be captured through a Schwarzchild or (to higher resolution) a Reissner–Nordstr\"{o}m metric. However, approximations used for our purposes, namely keeping terms to $\mathcal{O}(q)$ in the action and $\mathcal{O}(M^{-2})$ in Eq.~\eqref{eq:mass_shell}, indicate a resolution good to $\mathcal{O}(r^{-1})$ in our induced metric. We leave the derivation of the Schwarzchild metric as a topic for future work. 
\vspace{0.5em}

\section{Noncommutative Description}

Let us understand why the addition of a gauge in momentum space can lead to the long range/low energy description of gravity; this is due to the underlying noncommutative structure that emerges. 
We can readily show the structure in the first-quantized representation by investigating the equations of motions in Eq.~\eqref{eq:eoms}. To do so, we will consider an enlarged phase-space, described by variables $\xi_\Lambda \coloneqq(x_\mu,p_\nu)$, where capital Greek indices run over the 8 spacetime indices corresponding to the $(1+3)\oplus(1+3)$-dimensions. The worldline Lagrangian corresponding to Eq.~\eqref{eq:w_action} may then be written as $L_\text{W}=a^{\Lambda}\dot{\xi}_{\Lambda}-H(\xi)$, where
\begin{equation}
    a^\Lambda=\begin{pmatrix}p+qA\\bB\end{pmatrix}\,,\quad H(\xi)=T( p^{2}+ m^{2}c^{2})\,.
\end{equation}
Then the (Hamilton) equations of motion read
\begin{equation}
    \dot{\xi}_{\Lambda}=\{H,\xi_{\Lambda}\}=(\Omega^{-1})_{\Lambda \Gamma}\partial^{\Gamma}H\eqqcolon\Omega_{\Lambda \Gamma}\partial^{\Gamma}H\,.
\end{equation}
Correspondingly we have that $\Omega^{\Lambda \Gamma}\dot{\xi}_{\Gamma}=\partial^{\Lambda}H$, and $\Omega^{\Lambda \Gamma}=\partial^{\Lambda}a^{\Gamma}-\partial^{\Gamma}a^{\Lambda}$.
Here the Poisson brackets read $\{f,g\}=\Omega_{\Lambda \Gamma}\partial^{\Lambda}f\partial^{\Gamma}g$. One then need only find find the inverse of 
\begin{equation}\label{eq:Omega}
    \Omega^{\Lambda \Gamma}=\begin{pmatrix}qF & -\eta\\\eta & bS \end{pmatrix}
\end{equation}
to show the noncanonical nature. Let us first write the inverse of $\omega^{\mu\nu}$ given in Eq.~\eqref{eq:omega} as~\cite{PhysRevD.105.116014}
\begin{equation}
    (\omega^{-1})_{\mu\nu}\eqqcolon\omega_{\mu\nu}=\chi^{-1}\bigl(\eta_{\mu\nu}-qb\widetilde{F}_{\mu\rho}\widetilde{S}^\rho_{\hphantom{\rho}\nu}\bigr)\,,
\end{equation}
where $\chi=1-qb\,I_{SF}-(qb)^{2}I_{\tilde{F}F}I_{\tilde{S}S}$, and $\chi^2=-\mathrm{det}\omega_{\mu\nu}$. Here the Lorentz invariants are $I_{\tilde{F}F} =-\tfrac{1}{4}(\tfrac{1}{2}\epsilon_{\mu\nu\alpha\beta}F^{\mu\nu}F^{\alpha\beta}):=-\tfrac{1}{4}\widetilde{F}_{\mu\nu}F^{\mu\nu}$, $I_{\tilde{S}S} =-\frac{1}{4}\widetilde{S}_{\mu\nu}S^{\mu\nu}$, and $I_{SF} =\frac{1}{2}S_{\mu\nu}F^{\mu\nu}$. The above expression follows from Cayley-Hamilton's theorem, and can be confirmed using
\begin{equation}
    \widetilde{F}F=I_{\tilde{F}F}\eta\,,\quad\widetilde{S}S=I_{\tilde{S}S}\eta\,,\quad SF-\widetilde{F}\widetilde{S}=-I_{SF}\eta\,.
\end{equation}
Expressed in terms of $(\omega^{-1})_{\mu\nu}$ the inverse of Eq.~\eqref{eq:Omega} can be readily found as
\begin{equation}
    \Omega_{\Lambda \Gamma}=\begin{pmatrix}\omega^{-1}bS & \omega^{-1}\\-\omega^{-1}& qF\omega^{-1} \end{pmatrix}\,,
\end{equation}
from which one may read off the Poisson brackets as
\begin{align}
    \{x_{\mu},x_{\nu}\} &=(\omega^{-1}bS)_{\mu\nu}\,,\quad \{p_{\mu},p_{\nu}\}=(qF\omega^{-1})_{\mu\nu}\,,\nonumber\\
    \{x_{\mu},p_{\nu}\} &=\omega_{\mu\nu}\,.
    \label{eq:noncommutative}
\end{align}
We can confirm that the Berry-inspired gauge does indeed give rise to noncanonical Poisson brackets, and shares a similar form to the 3-dimensional Poisson brackets found in~\cite{doi:10.1142/S0217984906010573}. In the absence of the Berry-inspired gauge we see that the ordinary canonical Poisson brackets are recovered. The above deformation of the canonical Poisson brackets are similar to other theories under the umbrella of doubly, or deformed, special relativity~\cite{AMELINOCAMELIA2001255,*KOWALSKIGLIKMAN2001391,*doi:10.1142/S0218271802001330} 
that include for example Snyder~\cite{PhysRev.71.38}, $\kappa$-Minkowski~\cite{LUKIERSKI199590}, and Magueijo-Smolin~\cite{PhysRevD.67.044017} spacetimes in that there is momentum dependence in the noncanonical term. Such theories can be conveniently encompassed in the form~\cite{doi:10.1142/S0217732320501801} $\{x_{\mu},x_{\nu}\} =\psi_{\mu\nu}^{\hphantom{\mu\nu}\rho}(p)x_\rho$, $\{p_\mu,x_\nu\}=\Xi_{\mu\nu}(p)$, and $\{p_\mu,p_\nu\}=0$ for arbitrary $\psi$ and $\Xi$. A key difference of Eq.~\eqref{eq:noncommutative} to the above is the coupling to electromagnetism. 
One may also note that no $x_{\mu}$ proportional term is present in the right side of $\{x_{\mu}, x_{\nu}\}$ of Eq.~\eqref{eq:noncommutative}.

To understand the noncommutative structure better and from a second-quantized field perspective let us write down the corresponding quantum canonical commutation relation in coordinate space to lowest order in $b$ coupling, which is $[\hat{x}_{\mu},\hat{x}_{\nu}]=i\hbar bS_{\mu\nu}(p)$, illustrating the noncommutativity via the Berry-inspired gauge field strength, $S_{\mu\nu}$, as argued in~\cite{Guendelman:2022gue}. Such a relation follows from a ``covariant derivative"~\cite{doi:10.1126/science.1089408}, but supplementing the coordinates such that
$\hat{x}_{\mu}\coloneqq x_{\mu}+bB_{\mu}(p)$. A feature is that there are canonical phase-space variables obeying $[x_{\mu},p_{\nu}]=i\hbar\eta_{\mu\nu}$ and $[x_\mu,x_\nu]=[p_\mu,p_\nu]=0$. 
It is intriguing to depict the functional form of the new covariant derivative's as it stems from a simple U$_p(1)$ phase transformation.
Let us consider simply the gauge $A_\mu(x)$ under the rotation, or $A'_\mu=\exp(if(p))A_\mu(x)\exp(-if(p))$, where $p_\mu=-i\hbar\partial_\mu$ is understood to be an operator. Next, we assume a Fock-Schwinger gauge expansion for $A_\mu(x)$~\cite{SHIFMAN198013}. 
Then using the canonical commutation relations one can determine that the gauge field under a U$_p(1)$ rotation acquires the addition as $A_\mu'=A_\mu(x+\hbar \partial^p f)$.
\vspace{0.5em}

\section{Conclusions and Outlook}

Motivated by the appearance of a Berry phase in momentum space that appears ubiquitously in condensed matter, and has been argued to exist on the worldline for fermions~\cite{PhysRevD.105.116014}, we study a local and dynamical Abelian extension of the symmetry. Coupled with electromagnetism we find a gauge in momentum space and electromagnetic field strength dependent emergent metric appears from the resulting noncommutative d.o.f. To further illustrate the utility of a dynamical gauge in momentum space, we examine its classical solutions for a point-like particle that are Coulomb-like; then coupled with an electromagnetic Coulomb potential we argue such a dual Coulomb description predicts a Newtonian potential. 

We have examined a U$_p(1)$ gauge in momentum space in a physically opaque setting for complex scalars, and indeed the symmetry also exists for fermions. However, for the case of fermions, one may also argue a local and dynamical extension of the SO$(1,3)$ symmetry for spinors in momentum space; such a formulation to study a curved momentum space for fermions has been employed in~\cite{Franchino-Vinas:2022fkh}. Then, in contrast to the Abelian U$_p(1)$ group, for SO$(1,3)$ more complex, and topologically non-trivial, classical solutions may be present, such as is already the case for Berry's phase~\cite{PhysRevD.105.116014}. Then topological solutions involving a pure gauge in SO$(1,3)$, (such as are, e.g., the case for instantons, merons, etc. in SU$(2)$ Yang-Mills theory), would be suppressed by the Planck constant $\hbar$, as would emergent geometry, and perhaps help to explain the weakness of gravity as a quantum phenomena appearing at the classical level. This is also a subject of future work.
\vspace{0.5em}

\begin{acknowledgments}
The authors would like to thank Sebasti\'{a}n A. Franchino-Vi\~{n}as and Salvatore Mignemi for fruitful comments and discussions. P.C. acknowledges support from the Institute of Physics in Academia Sinica, Taiwan.
\end{acknowledgments}

\bibliography{references}

\begin{thebibliography}{63}%
\makeatletter
\providecommand \@ifxundefined [1]{%
 \@ifx{#1\undefined}
}%
\providecommand \@ifnum [1]{%
 \ifnum #1\expandafter \@firstoftwo
 \else \expandafter \@secondoftwo
 \fi
}%
\providecommand \@ifx [1]{%
 \ifx #1\expandafter \@firstoftwo
 \else \expandafter \@secondoftwo
 \fi
}%
\providecommand \natexlab [1]{#1}%
\providecommand \enquote  [1]{``#1''}%
\providecommand \bibnamefont  [1]{#1}%
\providecommand \bibfnamefont [1]{#1}%
\providecommand \citenamefont [1]{#1}%
\providecommand \href@noop [0]{\@secondoftwo}%
\providecommand \href [0]{\begingroup \@sanitize@url \@href}%
\providecommand \@href[1]{\@@startlink{#1}\@@href}%
\providecommand \@@href[1]{\endgroup#1\@@endlink}%
\providecommand \@sanitize@url [0]{\catcode `\\12\catcode `\$12\catcode
  `\&12\catcode `\#12\catcode `\^12\catcode `\_12\catcode `\%12\relax}%
\providecommand \@@startlink[1]{}%
\providecommand \@@endlink[0]{}%
\providecommand \url  [0]{\begingroup\@sanitize@url \@url }%
\providecommand \@url [1]{\endgroup\@href {#1}{\urlprefix }}%
\providecommand \urlprefix  [0]{URL }%
\providecommand \Eprint [0]{\href }%
\providecommand \doibase [0]{http://dx.doi.org/}%
\providecommand \selectlanguage [0]{\@gobble}%
\providecommand \bibinfo  [0]{\@secondoftwo}%
\providecommand \bibfield  [0]{\@secondoftwo}%
\providecommand \translation [1]{[#1]}%
\providecommand \BibitemOpen [0]{}%
\providecommand \bibitemStop [0]{}%
\providecommand \bibitemNoStop [0]{.\EOS\space}%
\providecommand \EOS [0]{\spacefactor3000\relax}%
\providecommand \BibitemShut  [1]{\csname bibitem#1\endcsname}%
\let\auto@bib@innerbib\@empty
\bibitem [{\citenamefont {Berry}(1987)}]{doi:10.1098/rspa.1987.0131}%
  \BibitemOpen
  \bibfield  {author} {\bibinfo {author} {\bibfnamefont {M.~V.}\ \bibnamefont
  {Berry}},\ }\href {\doibase 10.1098/rspa.1987.0131} {\bibfield  {journal}
  {\bibinfo  {journal} {Proceedings of the Royal Society of London. A.
  Mathematical and Physical Sciences}\ }\textbf {\bibinfo {volume} {414}},\
  \bibinfo {pages} {31} (\bibinfo {year} {1987})}\BibitemShut {NoStop}%
\bibitem [{\citenamefont {Thouless}\ \emph {et~al.}(1982)\citenamefont
  {Thouless}, \citenamefont {Kohmoto}, \citenamefont {Nightingale},\ and\
  \citenamefont {den Nijs}}]{Thouless:1982zz}%
  \BibitemOpen
  \bibfield  {author} {\bibinfo {author} {\bibfnamefont {D.~J.}\ \bibnamefont
  {Thouless}}, \bibinfo {author} {\bibfnamefont {M.}~\bibnamefont {Kohmoto}},
  \bibinfo {author} {\bibfnamefont {M.~P.}\ \bibnamefont {Nightingale}}, \ and\
  \bibinfo {author} {\bibfnamefont {M.}~\bibnamefont {den Nijs}},\ }\href
  {\doibase 10.1103/PhysRevLett.49.405} {\bibfield  {journal} {\bibinfo
  {journal} {Phys. Rev. Lett.}\ }\textbf {\bibinfo {volume} {49}},\ \bibinfo
  {pages} {405} (\bibinfo {year} {1982})}\BibitemShut {NoStop}%
\bibitem [{\citenamefont {Chang}\ and\ \citenamefont
  {Niu}(1995)}]{PhysRevLett.75.1348}%
  \BibitemOpen
  \bibfield  {author} {\bibinfo {author} {\bibfnamefont {M.-C.}\ \bibnamefont
  {Chang}}\ and\ \bibinfo {author} {\bibfnamefont {Q.}~\bibnamefont {Niu}},\
  }\href {\doibase 10.1103/PhysRevLett.75.1348} {\bibfield  {journal} {\bibinfo
   {journal} {Phys. Rev. Lett.}\ }\textbf {\bibinfo {volume} {75}},\ \bibinfo
  {pages} {1348} (\bibinfo {year} {1995})}\BibitemShut {NoStop}%
\bibitem [{\citenamefont {Chang}\ and\ \citenamefont
  {Niu}(1996)}]{PhysRevB.53.7010}%
  \BibitemOpen
  \bibfield  {author} {\bibinfo {author} {\bibfnamefont {M.-C.}\ \bibnamefont
  {Chang}}\ and\ \bibinfo {author} {\bibfnamefont {Q.}~\bibnamefont {Niu}},\
  }\href {\doibase 10.1103/PhysRevB.53.7010} {\bibfield  {journal} {\bibinfo
  {journal} {Phys. Rev. B}\ }\textbf {\bibinfo {volume} {53}},\ \bibinfo
  {pages} {7010} (\bibinfo {year} {1996})}\BibitemShut {NoStop}%
\bibitem [{\citenamefont {Karplus}\ and\ \citenamefont
  {Luttinger}(1954)}]{PhysRev.95.1154}%
  \BibitemOpen
  \bibfield  {author} {\bibinfo {author} {\bibfnamefont {R.}~\bibnamefont
  {Karplus}}\ and\ \bibinfo {author} {\bibfnamefont {J.~M.}\ \bibnamefont
  {Luttinger}},\ }\href {\doibase 10.1103/PhysRev.95.1154} {\bibfield
  {journal} {\bibinfo  {journal} {Phys. Rev.}\ }\textbf {\bibinfo {volume}
  {95}},\ \bibinfo {pages} {1154} (\bibinfo {year} {1954})}\BibitemShut
  {NoStop}%
\bibitem [{\citenamefont {Taguchi}\ \emph {et~al.}(2001)\citenamefont
  {Taguchi}, \citenamefont {Oohara}, \citenamefont {Yoshizawa}, \citenamefont
  {Nagaosa},\ and\ \citenamefont {Tokura}}]{doi:10.1126/science.1058161}%
  \BibitemOpen
  \bibfield  {author} {\bibinfo {author} {\bibfnamefont {Y.}~\bibnamefont
  {Taguchi}}, \bibinfo {author} {\bibfnamefont {Y.}~\bibnamefont {Oohara}},
  \bibinfo {author} {\bibfnamefont {H.}~\bibnamefont {Yoshizawa}}, \bibinfo
  {author} {\bibfnamefont {N.}~\bibnamefont {Nagaosa}}, \ and\ \bibinfo
  {author} {\bibfnamefont {Y.}~\bibnamefont {Tokura}},\ }\href {\doibase
  10.1126/science.1058161} {\bibfield  {journal} {\bibinfo  {journal}
  {Science}\ }\textbf {\bibinfo {volume} {291}},\ \bibinfo {pages} {2573}
  (\bibinfo {year} {2001})}\BibitemShut {NoStop}%
\bibitem [{\citenamefont {Jungwirth}\ \emph {et~al.}(2002)\citenamefont
  {Jungwirth}, \citenamefont {Niu},\ and\ \citenamefont
  {MacDonald}}]{PhysRevLett.88.207208}%
  \BibitemOpen
  \bibfield  {author} {\bibinfo {author} {\bibfnamefont {T.}~\bibnamefont
  {Jungwirth}}, \bibinfo {author} {\bibfnamefont {Q.}~\bibnamefont {Niu}}, \
  and\ \bibinfo {author} {\bibfnamefont {A.~H.}\ \bibnamefont {MacDonald}},\
  }\href {\doibase 10.1103/PhysRevLett.88.207208} {\bibfield  {journal}
  {\bibinfo  {journal} {Phys. Rev. Lett.}\ }\textbf {\bibinfo {volume} {88}},\
  \bibinfo {pages} {207208} (\bibinfo {year} {2002})}\BibitemShut {NoStop}%
\bibitem [{\citenamefont {Fang}\ \emph {et~al.}(2003)\citenamefont {Fang},
  \citenamefont {Nagaosa}, \citenamefont {Takahashi}, \citenamefont {Asamitsu},
  \citenamefont {Mathieu}, \citenamefont {Ogasawara}, \citenamefont {Yamada},
  \citenamefont {Kawasaki}, \citenamefont {Tokura},\ and\ \citenamefont
  {Terakura}}]{doi:10.1126/science.1089408}%
  \BibitemOpen
  \bibfield  {author} {\bibinfo {author} {\bibfnamefont {Z.}~\bibnamefont
  {Fang}}, \bibinfo {author} {\bibfnamefont {N.}~\bibnamefont {Nagaosa}},
  \bibinfo {author} {\bibfnamefont {K.~S.}\ \bibnamefont {Takahashi}}, \bibinfo
  {author} {\bibfnamefont {A.}~\bibnamefont {Asamitsu}}, \bibinfo {author}
  {\bibfnamefont {R.}~\bibnamefont {Mathieu}}, \bibinfo {author} {\bibfnamefont
  {T.}~\bibnamefont {Ogasawara}}, \bibinfo {author} {\bibfnamefont
  {H.}~\bibnamefont {Yamada}}, \bibinfo {author} {\bibfnamefont
  {M.}~\bibnamefont {Kawasaki}}, \bibinfo {author} {\bibfnamefont
  {Y.}~\bibnamefont {Tokura}}, \ and\ \bibinfo {author} {\bibfnamefont
  {K.}~\bibnamefont {Terakura}},\ }\href {\doibase 10.1126/science.1089408}
  {\bibfield  {journal} {\bibinfo  {journal} {Science}\ }\textbf {\bibinfo
  {volume} {302}},\ \bibinfo {pages} {92} (\bibinfo {year} {2003})}\BibitemShut
  {NoStop}%
\bibitem [{\citenamefont {Murakami}\ \emph {et~al.}(2003)\citenamefont
  {Murakami}, \citenamefont {Nagaosa},\ and\ \citenamefont
  {Zhang}}]{doi:10.1126/science.1087128}%
  \BibitemOpen
  \bibfield  {author} {\bibinfo {author} {\bibfnamefont {S.}~\bibnamefont
  {Murakami}}, \bibinfo {author} {\bibfnamefont {N.}~\bibnamefont {Nagaosa}}, \
  and\ \bibinfo {author} {\bibfnamefont {S.-C.}\ \bibnamefont {Zhang}},\ }\href
  {\doibase 10.1126/science.1087128} {\bibfield  {journal} {\bibinfo  {journal}
  {Science}\ }\textbf {\bibinfo {volume} {301}},\ \bibinfo {pages} {1348}
  (\bibinfo {year} {2003})}\BibitemShut {NoStop}%
\bibitem [{\citenamefont {Sinova}\ \emph {et~al.}(2004)\citenamefont {Sinova},
  \citenamefont {Culcer}, \citenamefont {Niu}, \citenamefont {Sinitsyn},
  \citenamefont {Jungwirth},\ and\ \citenamefont
  {MacDonald}}]{PhysRevLett.92.126603}%
  \BibitemOpen
  \bibfield  {author} {\bibinfo {author} {\bibfnamefont {J.}~\bibnamefont
  {Sinova}}, \bibinfo {author} {\bibfnamefont {D.}~\bibnamefont {Culcer}},
  \bibinfo {author} {\bibfnamefont {Q.}~\bibnamefont {Niu}}, \bibinfo {author}
  {\bibfnamefont {N.~A.}\ \bibnamefont {Sinitsyn}}, \bibinfo {author}
  {\bibfnamefont {T.}~\bibnamefont {Jungwirth}}, \ and\ \bibinfo {author}
  {\bibfnamefont {A.~H.}\ \bibnamefont {MacDonald}},\ }\href {\doibase
  10.1103/PhysRevLett.92.126603} {\bibfield  {journal} {\bibinfo  {journal}
  {Phys. Rev. Lett.}\ }\textbf {\bibinfo {volume} {92}},\ \bibinfo {pages}
  {126603} (\bibinfo {year} {2004})}\BibitemShut {NoStop}%
\bibitem [{\citenamefont {Adler}(1969)}]{PhysRev.177.2426}%
  \BibitemOpen
  \bibfield  {author} {\bibinfo {author} {\bibfnamefont {S.~L.}\ \bibnamefont
  {Adler}},\ }\href {\doibase 10.1103/PhysRev.177.2426} {\bibfield  {journal}
  {\bibinfo  {journal} {Phys. Rev.}\ }\textbf {\bibinfo {volume} {177}},\
  \bibinfo {pages} {2426} (\bibinfo {year} {1969})}\BibitemShut {NoStop}%
\bibitem [{\citenamefont {Bell}\ and\ \citenamefont {Jackiw}()}]{ref1}%
  \BibitemOpen
  \bibfield  {author} {\bibinfo {author} {\bibfnamefont {J.~S.}\ \bibnamefont
  {Bell}}\ and\ \bibinfo {author} {\bibfnamefont {R.}~\bibnamefont {Jackiw}},\
  }\href {\doibase 10.1007/BF02823296} {\bibfield  {journal} {\bibinfo
  {journal} {Il Nuovo Cimento A}\ }\textbf {\bibinfo {volume} {60}},\ \bibinfo
  {pages} {47}}\BibitemShut {NoStop}%
\bibitem [{\citenamefont {Huang}\ \emph {et~al.}(2015)\citenamefont {Huang},
  \citenamefont {Zhao}, \citenamefont {Long}, \citenamefont {Wang},
  \citenamefont {Chen}, \citenamefont {Yang}, \citenamefont {Liang},
  \citenamefont {Xue}, \citenamefont {Weng}, \citenamefont {Fang},
  \citenamefont {Dai},\ and\ \citenamefont {Chen}}]{PhysRevX.5.031023}%
  \BibitemOpen
  \bibfield  {author} {\bibinfo {author} {\bibfnamefont {X.}~\bibnamefont
  {Huang}}, \bibinfo {author} {\bibfnamefont {L.}~\bibnamefont {Zhao}},
  \bibinfo {author} {\bibfnamefont {Y.}~\bibnamefont {Long}}, \bibinfo {author}
  {\bibfnamefont {P.}~\bibnamefont {Wang}}, \bibinfo {author} {\bibfnamefont
  {D.}~\bibnamefont {Chen}}, \bibinfo {author} {\bibfnamefont {Z.}~\bibnamefont
  {Yang}}, \bibinfo {author} {\bibfnamefont {H.}~\bibnamefont {Liang}},
  \bibinfo {author} {\bibfnamefont {M.}~\bibnamefont {Xue}}, \bibinfo {author}
  {\bibfnamefont {H.}~\bibnamefont {Weng}}, \bibinfo {author} {\bibfnamefont
  {Z.}~\bibnamefont {Fang}}, \bibinfo {author} {\bibfnamefont {X.}~\bibnamefont
  {Dai}}, \ and\ \bibinfo {author} {\bibfnamefont {G.}~\bibnamefont {Chen}},\
  }\href {\doibase 10.1103/PhysRevX.5.031023} {\bibfield  {journal} {\bibinfo
  {journal} {Phys. Rev. X}\ }\textbf {\bibinfo {volume} {5}},\ \bibinfo {pages}
  {031023} (\bibinfo {year} {2015})}\BibitemShut {NoStop}%
\bibitem [{\citenamefont {Xiong}\ \emph {et~al.}(2015)\citenamefont {Xiong},
  \citenamefont {Kushwaha}, \citenamefont {Liang}, \citenamefont {Krizan},
  \citenamefont {Hirschberger}, \citenamefont {Wang}, \citenamefont {Cava},\
  and\ \citenamefont {Ong}}]{doi:10.1126/science.aac6089}%
  \BibitemOpen
  \bibfield  {author} {\bibinfo {author} {\bibfnamefont {J.}~\bibnamefont
  {Xiong}}, \bibinfo {author} {\bibfnamefont {S.~K.}\ \bibnamefont {Kushwaha}},
  \bibinfo {author} {\bibfnamefont {T.}~\bibnamefont {Liang}}, \bibinfo
  {author} {\bibfnamefont {J.~W.}\ \bibnamefont {Krizan}}, \bibinfo {author}
  {\bibfnamefont {M.}~\bibnamefont {Hirschberger}}, \bibinfo {author}
  {\bibfnamefont {W.}~\bibnamefont {Wang}}, \bibinfo {author} {\bibfnamefont
  {R.~J.}\ \bibnamefont {Cava}}, \ and\ \bibinfo {author} {\bibfnamefont
  {N.~P.}\ \bibnamefont {Ong}},\ }\href {\doibase 10.1126/science.aac6089}
  {\bibfield  {journal} {\bibinfo  {journal} {Science}\ }\textbf {\bibinfo
  {volume} {350}},\ \bibinfo {pages} {413} (\bibinfo {year}
  {2015})}\BibitemShut {NoStop}%
\bibitem [{\citenamefont {Stephanov}\ and\ \citenamefont
  {Yin}(2012)}]{PhysRevLett.109.162001}%
  \BibitemOpen
  \bibfield  {author} {\bibinfo {author} {\bibfnamefont {M.~A.}\ \bibnamefont
  {Stephanov}}\ and\ \bibinfo {author} {\bibfnamefont {Y.}~\bibnamefont
  {Yin}},\ }\href {\doibase 10.1103/PhysRevLett.109.162001} {\bibfield
  {journal} {\bibinfo  {journal} {Phys. Rev. Lett.}\ }\textbf {\bibinfo
  {volume} {109}},\ \bibinfo {pages} {162001} (\bibinfo {year}
  {2012})}\BibitemShut {NoStop}%
\bibitem [{\citenamefont {Duval}\ and\ \citenamefont
  {Horváthy}(2000)}]{DUVAL2000284}%
  \BibitemOpen
  \bibfield  {author} {\bibinfo {author} {\bibfnamefont {C.}~\bibnamefont
  {Duval}}\ and\ \bibinfo {author} {\bibfnamefont {P.}~\bibnamefont
  {Horváthy}},\ }\href {\doibase
  https://doi.org/10.1016/S0370-2693(00)00341-5} {\bibfield  {journal}
  {\bibinfo  {journal} {Physics Letters B}\ }\textbf {\bibinfo {volume}
  {479}},\ \bibinfo {pages} {284} (\bibinfo {year} {2000})}\BibitemShut
  {NoStop}%
\bibitem [{\citenamefont {Horvathy}\ \emph {et~al.}(2010)\citenamefont
  {Horvathy}, \citenamefont {Martina},\ and\ \citenamefont
  {Stichel}}]{Horvathy:2010wv}%
  \BibitemOpen
  \bibfield  {author} {\bibinfo {author} {\bibfnamefont {P.~A.}\ \bibnamefont
  {Horvathy}}, \bibinfo {author} {\bibfnamefont {L.}~\bibnamefont {Martina}}, \
  and\ \bibinfo {author} {\bibfnamefont {P.~C.}\ \bibnamefont {Stichel}},\
  }\href {\doibase 10.3842/SIGMA.2010.060} {\bibfield  {journal} {\bibinfo
  {journal} {SIGMA}\ }\textbf {\bibinfo {volume} {6}},\ \bibinfo {pages} {060}
  (\bibinfo {year} {2010})},\ \Eprint {http://arxiv.org/abs/1002.4772}
  {arXiv:1002.4772 [hep-th]} \BibitemShut {NoStop}%
\bibitem [{\citenamefont {Duval}\ and\ \citenamefont
  {Horváthy}(2001)}]{C_Duval_2001}%
  \BibitemOpen
  \bibfield  {author} {\bibinfo {author} {\bibfnamefont {C.}~\bibnamefont
  {Duval}}\ and\ \bibinfo {author} {\bibfnamefont {P.~A.}\ \bibnamefont
  {Horváthy}},\ }\href {\doibase 10.1088/0305-4470/34/47/314} {\bibfield
  {journal} {\bibinfo  {journal} {Journal of Physics A: Mathematical and
  General}\ }\textbf {\bibinfo {volume} {34}},\ \bibinfo {pages} {10097}
  (\bibinfo {year} {2001})}\BibitemShut {NoStop}%
\bibitem [{\citenamefont {B\'erard}\ and\ \citenamefont
  {Mohrbach}(2004)}]{PhysRevD.69.127701}%
  \BibitemOpen
  \bibfield  {author} {\bibinfo {author} {\bibfnamefont {A.}~\bibnamefont
  {B\'erard}}\ and\ \bibinfo {author} {\bibfnamefont {H.}~\bibnamefont
  {Mohrbach}},\ }\href {\doibase 10.1103/PhysRevD.69.127701} {\bibfield
  {journal} {\bibinfo  {journal} {Phys. Rev. D}\ }\textbf {\bibinfo {volume}
  {69}},\ \bibinfo {pages} {127701} (\bibinfo {year} {2004})}\BibitemShut
  {NoStop}%
\bibitem [{\citenamefont {Duval}\ \emph
  {et~al.}(2006{\natexlab{a}})\citenamefont {Duval}, \citenamefont
  {Horv\'{a}th}, \citenamefont {Horv\'{a}thy}, \citenamefont {Martina},\ and\
  \citenamefont {Stichel}}]{doi:10.1142/S0217984906010573}%
  \BibitemOpen
  \bibfield  {author} {\bibinfo {author} {\bibfnamefont {C.}~\bibnamefont
  {Duval}}, \bibinfo {author} {\bibfnamefont {Z.}~\bibnamefont {Horv\'{a}th}},
  \bibinfo {author} {\bibfnamefont {P.~A.}\ \bibnamefont {Horv\'{a}thy}},
  \bibinfo {author} {\bibfnamefont {L.}~\bibnamefont {Martina}}, \ and\
  \bibinfo {author} {\bibfnamefont {P.~C.}\ \bibnamefont {Stichel}},\ }\href
  {\doibase 10.1142/S0217984906010573} {\bibfield  {journal} {\bibinfo
  {journal} {Modern Physics Letters B}\ }\textbf {\bibinfo {volume} {20}},\
  \bibinfo {pages} {373} (\bibinfo {year} {2006}{\natexlab{a}})}\BibitemShut
  {NoStop}%
\bibitem [{\citenamefont {Douglas}\ and\ \citenamefont
  {Nekrasov}(2001)}]{RevModPhys.73.977}%
  \BibitemOpen
  \bibfield  {author} {\bibinfo {author} {\bibfnamefont {M.~R.}\ \bibnamefont
  {Douglas}}\ and\ \bibinfo {author} {\bibfnamefont {N.~A.}\ \bibnamefont
  {Nekrasov}},\ }\href {\doibase 10.1103/RevModPhys.73.977} {\bibfield
  {journal} {\bibinfo  {journal} {Rev. Mod. Phys.}\ }\textbf {\bibinfo {volume}
  {73}},\ \bibinfo {pages} {977} (\bibinfo {year} {2001})}\BibitemShut
  {NoStop}%
\bibitem [{\citenamefont {Szabo}(2003)}]{SZABO2003207}%
  \BibitemOpen
  \bibfield  {author} {\bibinfo {author} {\bibfnamefont {R.~J.}\ \bibnamefont
  {Szabo}},\ }\href {\doibase https://doi.org/10.1016/S0370-1573(03)00059-0}
  {\bibfield  {journal} {\bibinfo  {journal} {Physics Reports}\ }\textbf
  {\bibinfo {volume} {378}},\ \bibinfo {pages} {207} (\bibinfo {year}
  {2003})}\BibitemShut {NoStop}%
\bibitem [{\citenamefont {Snyder}(1947)}]{PhysRev.71.38}%
  \BibitemOpen
  \bibfield  {author} {\bibinfo {author} {\bibfnamefont {H.~S.}\ \bibnamefont
  {Snyder}},\ }\href {\doibase 10.1103/PhysRev.71.38} {\bibfield  {journal}
  {\bibinfo  {journal} {Phys. Rev.}\ }\textbf {\bibinfo {volume} {71}},\
  \bibinfo {pages} {38} (\bibinfo {year} {1947})}\BibitemShut {NoStop}%
\bibitem [{\citenamefont {Amelino-Camelia}(2001)}]{AMELINOCAMELIA2001255}%
  \BibitemOpen
  \bibfield  {author} {\bibinfo {author} {\bibfnamefont {G.}~\bibnamefont
  {Amelino-Camelia}},\ }\href {\doibase
  https://doi.org/10.1016/S0370-2693(01)00506-8} {\bibfield  {journal}
  {\bibinfo  {journal} {Physics Letters B}\ }\textbf {\bibinfo {volume}
  {510}},\ \bibinfo {pages} {255} (\bibinfo {year} {2001})}\BibitemShut
  {NoStop}%
\bibitem [{\citenamefont {Kowalski-Glikman}(2001)}]{KOWALSKIGLIKMAN2001391}%
  \BibitemOpen
  \bibfield  {author} {\bibinfo {author} {\bibfnamefont {J.}~\bibnamefont
  {Kowalski-Glikman}},\ }\href {\doibase
  https://doi.org/10.1016/S0375-9601(01)00465-0} {\bibfield  {journal}
  {\bibinfo  {journal} {Physics Letters A}\ }\textbf {\bibinfo {volume}
  {286}},\ \bibinfo {pages} {391} (\bibinfo {year} {2001})}\BibitemShut
  {NoStop}%
\bibitem [{\citenamefont
  {Amelino-Camelia}(2002)}]{doi:10.1142/S0218271802001330}%
  \BibitemOpen
  \bibfield  {author} {\bibinfo {author} {\bibfnamefont {G.}~\bibnamefont
  {Amelino-Camelia}},\ }\href {\doibase 10.1142/S0218271802001330} {\bibfield
  {journal} {\bibinfo  {journal} {International Journal of Modern Physics D}\
  }\textbf {\bibinfo {volume} {11}},\ \bibinfo {pages} {35} (\bibinfo {year}
  {2002})}\BibitemShut {NoStop}%
\bibitem [{\citenamefont {Lukierski}\ \emph {et~al.}(1995)\citenamefont
  {Lukierski}, \citenamefont {Ruegg},\ and\ \citenamefont
  {Zakrzewski}}]{LUKIERSKI199590}%
  \BibitemOpen
  \bibfield  {author} {\bibinfo {author} {\bibfnamefont {J.}~\bibnamefont
  {Lukierski}}, \bibinfo {author} {\bibfnamefont {H.}~\bibnamefont {Ruegg}}, \
  and\ \bibinfo {author} {\bibfnamefont {W.}~\bibnamefont {Zakrzewski}},\
  }\href {\doibase https://doi.org/10.1006/aphy.1995.1092} {\bibfield
  {journal} {\bibinfo  {journal} {Annals of Physics}\ }\textbf {\bibinfo
  {volume} {243}},\ \bibinfo {pages} {90} (\bibinfo {year} {1995})}\BibitemShut
  {NoStop}%
\bibitem [{\citenamefont {Magueijo}\ and\ \citenamefont
  {Smolin}(2003)}]{PhysRevD.67.044017}%
  \BibitemOpen
  \bibfield  {author} {\bibinfo {author} {\bibfnamefont {J.~a.}\ \bibnamefont
  {Magueijo}}\ and\ \bibinfo {author} {\bibfnamefont {L.}~\bibnamefont
  {Smolin}},\ }\href {\doibase 10.1103/PhysRevD.67.044017} {\bibfield
  {journal} {\bibinfo  {journal} {Phys. Rev. D}\ }\textbf {\bibinfo {volume}
  {67}},\ \bibinfo {pages} {044017} (\bibinfo {year} {2003})}\BibitemShut
  {NoStop}%
\bibitem [{\citenamefont {Szabo}(2010)}]{Szabo:2009tn}%
  \BibitemOpen
  \bibfield  {author} {\bibinfo {author} {\bibfnamefont {R.~J.}\ \bibnamefont
  {Szabo}},\ }\href {\doibase 10.1007/s10714-009-0897-4} {\bibfield  {journal}
  {\bibinfo  {journal} {Gen. Rel. Grav.}\ }\textbf {\bibinfo {volume} {42}},\
  \bibinfo {pages} {1} (\bibinfo {year} {2010})},\ \Eprint
  {http://arxiv.org/abs/0906.2913} {arXiv:0906.2913 [hep-th]} \BibitemShut
  {NoStop}%
\bibitem [{\citenamefont {Peierls}(1933)}]{peierls1933theorie}%
  \BibitemOpen
  \bibfield  {author} {\bibinfo {author} {\bibfnamefont {R.}~\bibnamefont
  {Peierls}},\ }\href@noop {} {\bibfield  {journal} {\bibinfo  {journal}
  {Zeitschrift f{\"u}r Physik}\ }\textbf {\bibinfo {volume} {80}},\ \bibinfo
  {pages} {763} (\bibinfo {year} {1933})}\BibitemShut {NoStop}%
\bibitem [{\citenamefont {Connes}\ \emph {et~al.}(1998)\citenamefont {Connes},
  \citenamefont {Douglas},\ and\ \citenamefont {Schwarz}}]{Alain_Connes_1998}%
  \BibitemOpen
  \bibfield  {author} {\bibinfo {author} {\bibfnamefont {A.}~\bibnamefont
  {Connes}}, \bibinfo {author} {\bibfnamefont {M.~R.}\ \bibnamefont {Douglas}},
  \ and\ \bibinfo {author} {\bibfnamefont {A.}~\bibnamefont {Schwarz}},\ }\href
  {\doibase 10.1088/1126-6708/1998/02/003} {\bibfield  {journal} {\bibinfo
  {journal} {Journal of High Energy Physics}\ }\textbf {\bibinfo {volume}
  {1998}},\ \bibinfo {pages} {003} (\bibinfo {year} {1998})}\BibitemShut
  {NoStop}%
\bibitem [{\citenamefont {Cortes}\ and\ \citenamefont
  {Plyushchay}(1996)}]{Cortes:1995wa}%
  \BibitemOpen
  \bibfield  {author} {\bibinfo {author} {\bibfnamefont {J.~L.}\ \bibnamefont
  {Cortes}}\ and\ \bibinfo {author} {\bibfnamefont {M.~S.}\ \bibnamefont
  {Plyushchay}},\ }\href {\doibase 10.1142/S0217751X96001590} {\bibfield
  {journal} {\bibinfo  {journal} {Int. J. Mod. Phys. A}\ }\textbf {\bibinfo
  {volume} {11}},\ \bibinfo {pages} {3331} (\bibinfo {year} {1996})},\ \Eprint
  {http://arxiv.org/abs/hep-th/9505117} {arXiv:hep-th/9505117} \BibitemShut
  {NoStop}%
\bibitem [{\citenamefont {del Olmo}\ and\ \citenamefont
  {Plyushchay}(2006)}]{delOlmo:2005md}%
  \BibitemOpen
  \bibfield  {author} {\bibinfo {author} {\bibfnamefont {M.~A.}\ \bibnamefont
  {del Olmo}}\ and\ \bibinfo {author} {\bibfnamefont {M.~S.}\ \bibnamefont
  {Plyushchay}},\ }\href {\doibase 10.1016/j.aop.2006.03.001} {\bibfield
  {journal} {\bibinfo  {journal} {Annals Phys.}\ }\textbf {\bibinfo {volume}
  {321}},\ \bibinfo {pages} {2830} (\bibinfo {year} {2006})},\ \Eprint
  {http://arxiv.org/abs/hep-th/0508020} {arXiv:hep-th/0508020} \BibitemShut
  {NoStop}%
\bibitem [{\citenamefont {Rivelles}(2003)}]{RIVELLES2003191}%
  \BibitemOpen
  \bibfield  {author} {\bibinfo {author} {\bibfnamefont {V.~O.}\ \bibnamefont
  {Rivelles}},\ }\href {\doibase https://doi.org/10.1016/S0370-2693(03)00271-5}
  {\bibfield  {journal} {\bibinfo  {journal} {Physics Letters B}\ }\textbf
  {\bibinfo {volume} {558}},\ \bibinfo {pages} {191} (\bibinfo {year}
  {2003})}\BibitemShut {NoStop}%
\bibitem [{\citenamefont {Yang}(2006)}]{doi:10.1142/S0217732306021682}%
  \BibitemOpen
  \bibfield  {author} {\bibinfo {author} {\bibfnamefont {H.~S.}\ \bibnamefont
  {Yang}},\ }\href {\doibase 10.1142/S0217732306021682} {\bibfield  {journal}
  {\bibinfo  {journal} {Modern Physics Letters A}\ }\textbf {\bibinfo {volume}
  {21}},\ \bibinfo {pages} {2637} (\bibinfo {year} {2006})}\BibitemShut
  {NoStop}%
\bibitem [{\citenamefont {Yang}(2009)}]{doi:10.1142/S0217751X0904587X}%
  \BibitemOpen
  \bibfield  {author} {\bibinfo {author} {\bibfnamefont {H.~S.}\ \bibnamefont
  {Yang}},\ }\href {\doibase 10.1142/S0217751X0904587X} {\bibfield  {journal}
  {\bibinfo  {journal} {International Journal of Modern Physics A}\ }\textbf
  {\bibinfo {volume} {24}},\ \bibinfo {pages} {4473} (\bibinfo {year}
  {2009})}\BibitemShut {NoStop}%
\bibitem [{\citenamefont {Seiberg}\ and\ \citenamefont
  {Witten}(1999)}]{Nathan_Seiberg_1999}%
  \BibitemOpen
  \bibfield  {author} {\bibinfo {author} {\bibfnamefont {N.}~\bibnamefont
  {Seiberg}}\ and\ \bibinfo {author} {\bibfnamefont {E.}~\bibnamefont
  {Witten}},\ }\href {\doibase 10.1088/1126-6708/1999/09/032} {\bibfield
  {journal} {\bibinfo  {journal} {Journal of High Energy Physics}\ }\textbf
  {\bibinfo {volume} {1999}},\ \bibinfo {pages} {032} (\bibinfo {year}
  {1999})}\BibitemShut {NoStop}%
\bibitem [{\citenamefont {Steinacker}(2007)}]{Harold_Steinacker_2007}%
  \BibitemOpen
  \bibfield  {author} {\bibinfo {author} {\bibfnamefont {H.}~\bibnamefont
  {Steinacker}},\ }\href {\doibase 10.1088/1126-6708/2007/12/049} {\bibfield
  {journal} {\bibinfo  {journal} {Journal of High Energy Physics}\ }\textbf
  {\bibinfo {volume} {2007}},\ \bibinfo {pages} {049} (\bibinfo {year}
  {2007})}\BibitemShut {NoStop}%
\bibitem [{\citenamefont {Steinacker}(2009)}]{STEINACKER20091}%
  \BibitemOpen
  \bibfield  {author} {\bibinfo {author} {\bibfnamefont {H.}~\bibnamefont
  {Steinacker}},\ }\href {\doibase
  https://doi.org/10.1016/j.nuclphysb.2008.10.014} {\bibfield  {journal}
  {\bibinfo  {journal} {Nuclear Physics B}\ }\textbf {\bibinfo {volume}
  {810}},\ \bibinfo {pages} {1} (\bibinfo {year} {2009})}\BibitemShut {NoStop}%
\bibitem [{\citenamefont {Majid}(1995)}]{majid_1995}%
  \BibitemOpen
  \bibfield  {author} {\bibinfo {author} {\bibfnamefont {S.}~\bibnamefont
  {Majid}},\ }\href {\doibase 10.1017/CBO9780511613104} {\emph {\bibinfo
  {title} {Foundations of Quantum Group Theory}}}\ (\bibinfo  {publisher}
  {Cambridge University Press},\ \bibinfo {year} {1995})\BibitemShut {NoStop}%
\bibitem [{\citenamefont {Stone}\ \emph {et~al.}(2015)\citenamefont {Stone},
  \citenamefont {Dwivedi},\ and\ \citenamefont {Zhou}}]{PhysRevD.91.025004}%
  \BibitemOpen
  \bibfield  {author} {\bibinfo {author} {\bibfnamefont {M.}~\bibnamefont
  {Stone}}, \bibinfo {author} {\bibfnamefont {V.}~\bibnamefont {Dwivedi}}, \
  and\ \bibinfo {author} {\bibfnamefont {T.}~\bibnamefont {Zhou}},\ }\href
  {\doibase 10.1103/PhysRevD.91.025004} {\bibfield  {journal} {\bibinfo
  {journal} {Phys. Rev. D}\ }\textbf {\bibinfo {volume} {91}},\ \bibinfo
  {pages} {025004} (\bibinfo {year} {2015})}\BibitemShut {NoStop}%
\bibitem [{\citenamefont {Guendelman}\ and\ \citenamefont
  {Singleton}(2022)}]{Guendelman:2022gue}%
  \BibitemOpen
  \bibfield  {author} {\bibinfo {author} {\bibfnamefont {E.}~\bibnamefont
  {Guendelman}}\ and\ \bibinfo {author} {\bibfnamefont {D.}~\bibnamefont
  {Singleton}},\ }\href@noop {} {\  (\bibinfo {year} {2022})},\ \Eprint
  {http://arxiv.org/abs/2206.02638} {arXiv:2206.02638 [quant-ph]} \BibitemShut
  {NoStop}%
\bibitem [{\citenamefont {Guendelman}\ and\ \citenamefont
  {Wagner}(2022)}]{Guendelman:2022ruu}%
  \BibitemOpen
  \bibfield  {author} {\bibinfo {author} {\bibfnamefont {E.}~\bibnamefont
  {Guendelman}}\ and\ \bibinfo {author} {\bibfnamefont {F.}~\bibnamefont
  {Wagner}},\ }\href@noop {} {\  (\bibinfo {year} {2022})},\ \Eprint
  {http://arxiv.org/abs/2208.00409} {arXiv:2208.00409 [gr-qc]} \BibitemShut
  {NoStop}%
\bibitem [{\citenamefont {Migdal}(1986)}]{MIGDAL1986594}%
  \BibitemOpen
  \bibfield  {author} {\bibinfo {author} {\bibfnamefont {A.}~\bibnamefont
  {Migdal}},\ }\href {\doibase https://doi.org/10.1016/0550-3213(86)90331-7}
  {\bibfield  {journal} {\bibinfo  {journal} {Nuclear Physics B}\ }\textbf
  {\bibinfo {volume} {265}},\ \bibinfo {pages} {594} (\bibinfo {year}
  {1986})}\BibitemShut {NoStop}%
\bibitem [{\citenamefont {Bonezzi}\ \emph {et~al.}(2012)\citenamefont
  {Bonezzi}, \citenamefont {Corradini}, \citenamefont {Viñas},\ and\
  \citenamefont {Pisani}}]{Bonezzi_2012}%
  \BibitemOpen
  \bibfield  {author} {\bibinfo {author} {\bibfnamefont {R.}~\bibnamefont
  {Bonezzi}}, \bibinfo {author} {\bibfnamefont {O.}~\bibnamefont {Corradini}},
  \bibinfo {author} {\bibfnamefont {S.~A.~F.}\ \bibnamefont {Viñas}}, \ and\
  \bibinfo {author} {\bibfnamefont {P.~A.~G.}\ \bibnamefont {Pisani}},\ }\href
  {\doibase 10.1088/1751-8113/45/40/405401} {\bibfield  {journal} {\bibinfo
  {journal} {Journal of Physics A: Mathematical and Theoretical}\ }\textbf
  {\bibinfo {volume} {45}},\ \bibinfo {pages} {405401} (\bibinfo {year}
  {2012})}\BibitemShut {NoStop}%
\bibitem [{\citenamefont {Franchino-Vi\~nas}\ and\ \citenamefont
  {Mignemi}(2018)}]{PhysRevD.98.065010}%
  \BibitemOpen
  \bibfield  {author} {\bibinfo {author} {\bibfnamefont {S.~A.}\ \bibnamefont
  {Franchino-Vi\~nas}}\ and\ \bibinfo {author} {\bibfnamefont {S.}~\bibnamefont
  {Mignemi}},\ }\href {\doibase 10.1103/PhysRevD.98.065010} {\bibfield
  {journal} {\bibinfo  {journal} {Phys. Rev. D}\ }\textbf {\bibinfo {volume}
  {98}},\ \bibinfo {pages} {065010} (\bibinfo {year} {2018})}\BibitemShut
  {NoStop}%
\bibitem [{\citenamefont {Copinger}\ and\ \citenamefont
  {Pu}(2022)}]{PhysRevD.105.116014}%
  \BibitemOpen
  \bibfield  {author} {\bibinfo {author} {\bibfnamefont {P.}~\bibnamefont
  {Copinger}}\ and\ \bibinfo {author} {\bibfnamefont {S.}~\bibnamefont {Pu}},\
  }\href {\doibase 10.1103/PhysRevD.105.116014} {\bibfield  {journal} {\bibinfo
   {journal} {Phys. Rev. D}\ }\textbf {\bibinfo {volume} {105}},\ \bibinfo
  {pages} {116014} (\bibinfo {year} {2022})}\BibitemShut {NoStop}%
\bibitem [{\citenamefont {Stone}\ and\ \citenamefont
  {Dwivedi}(2013)}]{PhysRevD.88.045012}%
  \BibitemOpen
  \bibfield  {author} {\bibinfo {author} {\bibfnamefont {M.}~\bibnamefont
  {Stone}}\ and\ \bibinfo {author} {\bibfnamefont {V.}~\bibnamefont
  {Dwivedi}},\ }\href {\doibase 10.1103/PhysRevD.88.045012} {\bibfield
  {journal} {\bibinfo  {journal} {Phys. Rev. D}\ }\textbf {\bibinfo {volume}
  {88}},\ \bibinfo {pages} {045012} (\bibinfo {year} {2013})}\BibitemShut
  {NoStop}%
\bibitem [{\citenamefont {Chen}\ \emph {et~al.}(2014)\citenamefont {Chen},
  \citenamefont {Pang}, \citenamefont {Pu},\ and\ \citenamefont
  {Wang}}]{PhysRevD.89.094003}%
  \BibitemOpen
  \bibfield  {author} {\bibinfo {author} {\bibfnamefont {J.-W.}\ \bibnamefont
  {Chen}}, \bibinfo {author} {\bibfnamefont {J.-y.}\ \bibnamefont {Pang}},
  \bibinfo {author} {\bibfnamefont {S.}~\bibnamefont {Pu}}, \ and\ \bibinfo
  {author} {\bibfnamefont {Q.}~\bibnamefont {Wang}},\ }\href {\doibase
  10.1103/PhysRevD.89.094003} {\bibfield  {journal} {\bibinfo  {journal} {Phys.
  Rev. D}\ }\textbf {\bibinfo {volume} {89}},\ \bibinfo {pages} {094003}
  (\bibinfo {year} {2014})}\BibitemShut {NoStop}%
\bibitem [{\citenamefont {Dwivedi}\ and\ \citenamefont
  {Stone}(2013)}]{Dwivedi_2014}%
  \BibitemOpen
  \bibfield  {author} {\bibinfo {author} {\bibfnamefont {V.}~\bibnamefont
  {Dwivedi}}\ and\ \bibinfo {author} {\bibfnamefont {M.}~\bibnamefont
  {Stone}},\ }\href {\doibase 10.1088/1751-8113/47/2/025401} {\bibfield
  {journal} {\bibinfo  {journal} {Journal of Physics A: Mathematical and
  Theoretical}\ }\textbf {\bibinfo {volume} {47}},\ \bibinfo {pages} {025401}
  (\bibinfo {year} {2013})}\BibitemShut {NoStop}%
\bibitem [{\citenamefont {Karanikas}\ \emph {et~al.}(1992)\citenamefont
  {Karanikas}, \citenamefont {Ktorides},\ and\ \citenamefont
  {Stefanis}}]{KARANIKAS1992176}%
  \BibitemOpen
  \bibfield  {author} {\bibinfo {author} {\bibfnamefont {A.}~\bibnamefont
  {Karanikas}}, \bibinfo {author} {\bibfnamefont {C.}~\bibnamefont {Ktorides}},
  \ and\ \bibinfo {author} {\bibfnamefont {N.}~\bibnamefont {Stefanis}},\
  }\href {\doibase https://doi.org/10.1016/0370-2693(92)91381-I} {\bibfield
  {journal} {\bibinfo  {journal} {Physics Letters B}\ }\textbf {\bibinfo
  {volume} {289}},\ \bibinfo {pages} {176} (\bibinfo {year}
  {1992})}\BibitemShut {NoStop}%
\bibitem [{\citenamefont {Franchino-Vi\~nas}\ and\ \citenamefont
  {Gies}(2019)}]{PhysRevD.100.105020}%
  \BibitemOpen
  \bibfield  {author} {\bibinfo {author} {\bibfnamefont {S.}~\bibnamefont
  {Franchino-Vi\~nas}}\ and\ \bibinfo {author} {\bibfnamefont {H.}~\bibnamefont
  {Gies}},\ }\href {\doibase 10.1103/PhysRevD.100.105020} {\bibfield  {journal}
  {\bibinfo  {journal} {Phys. Rev. D}\ }\textbf {\bibinfo {volume} {100}},\
  \bibinfo {pages} {105020} (\bibinfo {year} {2019})}\BibitemShut {NoStop}%
\bibitem [{\citenamefont {Xiao}\ \emph {et~al.}(2005)\citenamefont {Xiao},
  \citenamefont {Shi},\ and\ \citenamefont {Niu}}]{PhysRevLett.95.137204}%
  \BibitemOpen
  \bibfield  {author} {\bibinfo {author} {\bibfnamefont {D.}~\bibnamefont
  {Xiao}}, \bibinfo {author} {\bibfnamefont {J.}~\bibnamefont {Shi}}, \ and\
  \bibinfo {author} {\bibfnamefont {Q.}~\bibnamefont {Niu}},\ }\href {\doibase
  10.1103/PhysRevLett.95.137204} {\bibfield  {journal} {\bibinfo  {journal}
  {Phys. Rev. Lett.}\ }\textbf {\bibinfo {volume} {95}},\ \bibinfo {pages}
  {137204} (\bibinfo {year} {2005})}\BibitemShut {NoStop}%
\bibitem [{\citenamefont {Duval}\ \emph
  {et~al.}(2006{\natexlab{b}})\citenamefont {Duval}, \citenamefont {Horv\'ath},
  \citenamefont {Horv\'athy}, \citenamefont {Martina},\ and\ \citenamefont
  {Stichel}}]{PhysRevLett.96.099701}%
  \BibitemOpen
  \bibfield  {author} {\bibinfo {author} {\bibfnamefont {C.}~\bibnamefont
  {Duval}}, \bibinfo {author} {\bibfnamefont {Z.}~\bibnamefont {Horv\'ath}},
  \bibinfo {author} {\bibfnamefont {P.~A.}\ \bibnamefont {Horv\'athy}},
  \bibinfo {author} {\bibfnamefont {L.}~\bibnamefont {Martina}}, \ and\
  \bibinfo {author} {\bibfnamefont {P.~C.}\ \bibnamefont {Stichel}},\ }\href
  {\doibase 10.1103/PhysRevLett.96.099701} {\bibfield  {journal} {\bibinfo
  {journal} {Phys. Rev. Lett.}\ }\textbf {\bibinfo {volume} {96}},\ \bibinfo
  {pages} {099701} (\bibinfo {year} {2006}{\natexlab{b}})}\BibitemShut
  {NoStop}%
\bibitem [{\citenamefont {Jackiw}(1993)}]{Jackiw:1993in}%
  \BibitemOpen
  \bibfield  {author} {\bibinfo {author} {\bibfnamefont {R.}~\bibnamefont
  {Jackiw}},\ }in\ \href@noop {} {\emph {\bibinfo {booktitle} {{2nd Workshop on
  Constraint Theory and Quantization Methods}}}}\ (\bibinfo {year} {1993})\
  pp.\ \bibinfo {pages} {367--381},\ \Eprint
  {http://arxiv.org/abs/hep-th/9306075} {arXiv:hep-th/9306075} \BibitemShut
  {NoStop}%
\bibitem [{\citenamefont {Mignemi}\ and\ \citenamefont
  {Štrajn}(2016)}]{MIGNEMI20161714}%
  \BibitemOpen
  \bibfield  {author} {\bibinfo {author} {\bibfnamefont {S.}~\bibnamefont
  {Mignemi}}\ and\ \bibinfo {author} {\bibfnamefont {R.}~\bibnamefont
  {Štrajn}},\ }\href {\doibase https://doi.org/10.1016/j.physleta.2016.03.005}
  {\bibfield  {journal} {\bibinfo  {journal} {Physics Letters A}\ }\textbf
  {\bibinfo {volume} {380}},\ \bibinfo {pages} {1714} (\bibinfo {year}
  {2016})}\BibitemShut {NoStop}%
\bibitem [{\citenamefont {Balasin}\ \emph {et~al.}(2014)\citenamefont
  {Balasin}, \citenamefont {Blaschke}, \citenamefont {Gieres},\ and\
  \citenamefont {Schweda}}]{Balasin:2014dma}%
  \BibitemOpen
  \bibfield  {author} {\bibinfo {author} {\bibfnamefont {H.}~\bibnamefont
  {Balasin}}, \bibinfo {author} {\bibfnamefont {D.~N.}\ \bibnamefont
  {Blaschke}}, \bibinfo {author} {\bibfnamefont {F.}~\bibnamefont {Gieres}}, \
  and\ \bibinfo {author} {\bibfnamefont {M.}~\bibnamefont {Schweda}},\ }\href
  {\doibase 10.3842/SIGMA.2014.099} {\bibfield  {journal} {\bibinfo  {journal}
  {SIGMA}\ }\textbf {\bibinfo {volume} {10}},\ \bibinfo {pages} {099} (\bibinfo
  {year} {2014})},\ \Eprint {http://arxiv.org/abs/1403.0255} {arXiv:1403.0255
  [hep-th]} \BibitemShut {NoStop}%
\bibitem [{\citenamefont {Pavšič}(1987)}]{PAVSIC1987327}%
  \BibitemOpen
  \bibfield  {author} {\bibinfo {author} {\bibfnamefont {M.}~\bibnamefont
  {Pavšič}},\ }\href {\doibase https://doi.org/10.1016/0370-2693(87)90393-5}
  {\bibfield  {journal} {\bibinfo  {journal} {Physics Letters B}\ }\textbf
  {\bibinfo {volume} {197}},\ \bibinfo {pages} {327} (\bibinfo {year}
  {1987})}\BibitemShut {NoStop}%
\bibitem [{\citenamefont {Bastianelli}\ and\ \citenamefont
  {Zirotti}(2002)}]{BASTIANELLI2002372}%
  \BibitemOpen
  \bibfield  {author} {\bibinfo {author} {\bibfnamefont {F.}~\bibnamefont
  {Bastianelli}}\ and\ \bibinfo {author} {\bibfnamefont {A.}~\bibnamefont
  {Zirotti}},\ }\href {\doibase https://doi.org/10.1016/S0550-3213(02)00683-1}
  {\bibfield  {journal} {\bibinfo  {journal} {Nuclear Physics B}\ }\textbf
  {\bibinfo {volume} {642}},\ \bibinfo {pages} {372} (\bibinfo {year}
  {2002})}\BibitemShut {NoStop}%
\bibitem [{\citenamefont {Carroll}(2019)}]{carroll_2019}%
  \BibitemOpen
  \bibfield  {author} {\bibinfo {author} {\bibfnamefont {S.~M.}\ \bibnamefont
  {Carroll}},\ }\href {\doibase 10.1017/9781108770385} {\emph {\bibinfo {title}
  {Spacetime and Geometry: An Introduction to General Relativity}}}\ (\bibinfo
  {publisher} {Cambridge University Press},\ \bibinfo {year}
  {2019})\BibitemShut {NoStop}%
\bibitem [{\citenamefont {Mignemi}\ and\ \citenamefont
  {Rosati}(2020)}]{doi:10.1142/S0217732320501801}%
  \BibitemOpen
  \bibfield  {author} {\bibinfo {author} {\bibfnamefont {S.}~\bibnamefont
  {Mignemi}}\ and\ \bibinfo {author} {\bibfnamefont {G.}~\bibnamefont
  {Rosati}},\ }\href {\doibase 10.1142/S0217732320501801} {\bibfield  {journal}
  {\bibinfo  {journal} {Modern Physics Letters A}\ }\textbf {\bibinfo {volume}
  {35}},\ \bibinfo {pages} {2050180} (\bibinfo {year} {2020})}\BibitemShut
  {NoStop}%
\bibitem [{\citenamefont {Shifman}(1980)}]{SHIFMAN198013}%
  \BibitemOpen
  \bibfield  {author} {\bibinfo {author} {\bibfnamefont {M.~A.}\ \bibnamefont
  {Shifman}},\ }\href {\doibase https://doi.org/10.1016/0550-3213(80)90440-X}
  {\bibfield  {journal} {\bibinfo  {journal} {Nuclear Physics B}\ }\textbf
  {\bibinfo {volume} {173}},\ \bibinfo {pages} {13} (\bibinfo {year}
  {1980})}\BibitemShut {NoStop}%
\bibitem [{\citenamefont {Franchino-Vi\~nas}\ and\ \citenamefont
  {Relancio}(2022)}]{Franchino-Vinas:2022fkh}%
  \BibitemOpen
  \bibfield  {author} {\bibinfo {author} {\bibfnamefont {S.~A.}\ \bibnamefont
  {Franchino-Vi\~nas}}\ and\ \bibinfo {author} {\bibfnamefont {J.~J.}\
  \bibnamefont {Relancio}},\ }\href@noop {} {\  (\bibinfo {year} {2022})},\
  \Eprint {http://arxiv.org/abs/2203.12286} {arXiv:2203.12286 [hep-th]}
  \BibitemShut {NoStop}%
\end{thebibliography}%
\bibliographystyle{apsrev4-1}
\end{document}